\documentclass[%
 aip,
 apl,
 amsmath,amssymb,
 reprint,%
]{revtex4-1}

\usepackage{graphicx}
\usepackage{dcolumn}
\usepackage{bm}
\usepackage{subfigure}
\usepackage[utf8]{inputenc}
\usepackage[T1]{fontenc}
\usepackage{color}
\usepackage{mathptmx}

\begin{document}
\title{Designing 24-hour Electrical Power Generator: Thermoradiative Device for Harvesting Energy from Sun and Outer Space}

\author{Xin Zhang}
\email{zhangxin0605@jiangnan.edu.cn}
\author{Guofeng Yang}
\author{Mengqi Yan}
\affiliation{School of Science, Jiangnan University, Wuxi 214122, China}

\author{Lay Kee Ang}
\author{Yee Sin Ang} 
\email{yeesin_ang@sutd.edu.sg}
\affiliation{Science, Mathematics and Technology (SMT), Singapore University of Technology and Design, Singapore 487372}

\begin{abstract}
Energy harvesting from sun and outer space using thermoradiative devices (TRD), despite being promising renewable energy sources, are limited only to daytime and nighttime period, respectively. A system with 24-hour continuous energy generation remains an open question thus far. Here, we propose a TRD-based power generator that harvests solar energy via concentrated solar irradiation during daytime and via thermal infrared emission towards the outer space at nighttime, thus achieving the much sought-after 24-hour electrical power generation. We develop a rigorous thermodynamical model to investigate the performance characteristics, parametric optimum design, and the role of various energy loss mechanisms. Our model predicts that the TRD-based system yields a peak efficiency of 12.6\% at daytime and a maximum power density of 10.8 Wm$^{-2}$ at nighttime, thus significantly outperforming the state-of-art record-setting thermoelectric generator. These findings reveal the potential of TRD towards 24-hour electricity generation and future renewable energy technology.
\end{abstract}

\maketitle

Driven by the imminence of energy crisis, climate change and environmental pollution, the search of renewable energy sources and green energy haversting technologies have become some of the most important research centerpiece since the past century \cite{chu2017path,shindell2019climate,bagdadee2020electrical}. While solar power systems have offered a wide variety of electricity generation approaches, including photovoltaics and solar thermal power systems \cite{wu2020two,qu2019spectral}, the ability of generating electricity during both the daytime and nighttime with no necessity of energy storage remains an open question today. 
The lack of nighttime performance of a daytime energy haversting system, or the opposite case, necessitates the need for costly supplemental energy storage systems and/or grid connections to other nonrenewable energy sources \cite{hoppmann2014economic,xu2015application}, such as fossil fuels. 
Developing renewable energy harvesting technologies capable of continuously operating throughout 24 hours is strongly desirable, especially for supplying electricity to remote rural areas where access to electricity is challenging \cite{cabraal2005productive}.

Serving as a omnipresent resource to almost all parts of Earth, the outer space is an easily accessible thermodynamic cold sink, whose role in renewable energy applications has been largely ignored until very recently. 
Energy harvesting utilizing the coldness of the outer space have been extensively explored as a new route towards renewable energy future \cite{yin2020terrestrial,liu2020recent}, for example, 
the concept of electric power generation using radiative cooling and thermoelectric generators (TEGs) \cite{raman2019generating,fan2020maximal,zhao2020modeling}. 
Such radiative cooling system utilizes the outer space as a cold reservoir which allows the infrared thermal radiation from Earth’s surface to be dissipated through the atmospheric transparency window of wavelength, 8-13 $\mu$m. 
This concept of radiative cooling has been experimentally demonstrated \cite{raman2019generating} via the coupling between the cold side of the TEG composed of a near-black surface that radiates thermal radiation to outer space and the hot side heated by ambient air. 
Similarly, all-day electricity generation using radiative cooling and TEGs has also been actively explored \cite{chen2019simultaneously,tian2020harvesting,ao2020spectral,xia202024,liu220model,ishii2020radiative}. While these recent experimental efforts concretely demonstrated the potential usefulness of energy harvesting from space, their energy conversion performance is still limited by the relatively low figure-of-merit.

One feasible alternative to TEGs is to employ thermoradiative devices (TRDs) \cite{strandberg2015theoretical,fernandez2018endoreversible,zhang2019design,fernandez2019theoretical} that operates based on the thermoradiative effect of a semiconductor photodiode. 
Contrary to the operation of photovoltaics under positive sunlight illumination, TRD is an emerging heat-to-electricity technique that radiates a fraction of above-bandgap thermal radiation to the cold environment when operates at higher temperatures. 
Electrical power generation is achieved via photon exchange \cite{strandberg2015theoretical}. Experimentally, a thermoradiative semiconductor photodiode based on HgCdZnTe generates up to 1 pW of electrical power, which is insufficient for practical applications\cite{santhanam2016thermal}. 
To further improve the ability of power generation beyond the pW scale, taking the advantages of the naturally-existed temperature difference between the Sun, the environment, and the outer space to generate larger temperature difference and hence increases the output electrical power could be a feasible approach \cite{buddhiraju2018thermodynamic,deppe2019nighttime,li2020thermodynamic}. 
However, a number of open questions remain unaddressed so far:
(i) Can the concentrated solar thermal power and the thermoradiative effect be synnergized for harvesting energy from both sun and the outer space?
(ii) How to construct a rigorous computational model to understand the performance of such system in the presence of realistic optical, electric, and thermal losses? 
(iii) Which loss mechanism is the most dominant in limiting the system performance? 
(iv) What is the thermodynamic bounds of the system performance, and the optimal design strategies that can be applied to mitigate the effects that negatively impact the system performance? 

In this paper, we addressed the above questions by computationally designing a energy conversion system that couples a nighttime thermoradiative system with a daytime concentrated solar thermal power to produce electricity continuously throughout 24 hours. 
Such system is capable of providing small-scale, distributed renewable power generation for 24 hours a day and its performance is uniquely enabled by the optical coupling among Sun, outer space, and Earth's ambient environment. A peak efficiency of 12.6\% under 10 suns during daytime and a maximum power density of 10.8 Wm$^{-2}$ at nighttime room temperature are achieved without relying on a energy storage system nor active power input. Such performance significantly outperforms the TEG-based energy harvesting system, thus demonstrating TRD as a viable alternative route towards off-grid and battery-free renewable energy source.

Figure 1(a) shows an improved design of all-day thermoradiative system configuration that employs a three-dimensional structured graphene metamaterial (3D-SGM) solar absorber, a thermoradiative device (TRD) based on a HgCdTe p-n junction, and an Ag back surface reflector (BSR) under a parallel-plate geometry. The Sun ($\sim$6000K), the environment near the Earth’s surface ($\sim$300 K), and the outer space ($\sim$3 K) are three separate locations with a large temperature difference. 
Radiation heat transfer links the three thermodynamic resources through solar heating and thermal infrared radiation. 
During the daytime, the easily accessible sunlight serves as a heat source, and the ambient air acts as a cold sink to establish a consider temperature difference ($\sim$5700 K). 
An optical reflector allows the concentrated sunlight to impinge on the solar absorber, which convert entire solar spectrum into heat. 
The TRD absorbs a portion of the input thermal energy and operates at the temperature range of 500-800 K, generating electricity via photon exchange with the cold environment. 
At night, employing the warm environment as the heat reservoir and the outer space as the cold sink, the TRD radiates photons into the cold space through the atmospheric transparency window to generate electric power. 
A spectrum-selective mirror located on the right side reflects the infrared thermal emission, meanwhile, is transparent for other wavelength range so that the non-infrared radiation from the TRD can be emitted into the environment. The solar absorber is modeled based on a representative design of 3D-SGM with period 0.8 $\mu$m, trench depth 1 $\mu$m, hole width 0.59 $\mu$m, and film thickness 30 nm as reported in Ref. \cite{lin2020structured}. Importantly, 3D-SGM absorber exhibits many unique features, such as the superior selective absorption for solar spectrum, flexible tunability of wavelength-selective absorption, high thermal stability, and excellent opto-thermal performance, which are particularly beneficial for designing high-performance solar-thermal energy harvesting device.

\begin{figure}[t]
\centering
\includegraphics[height=8.758cm]{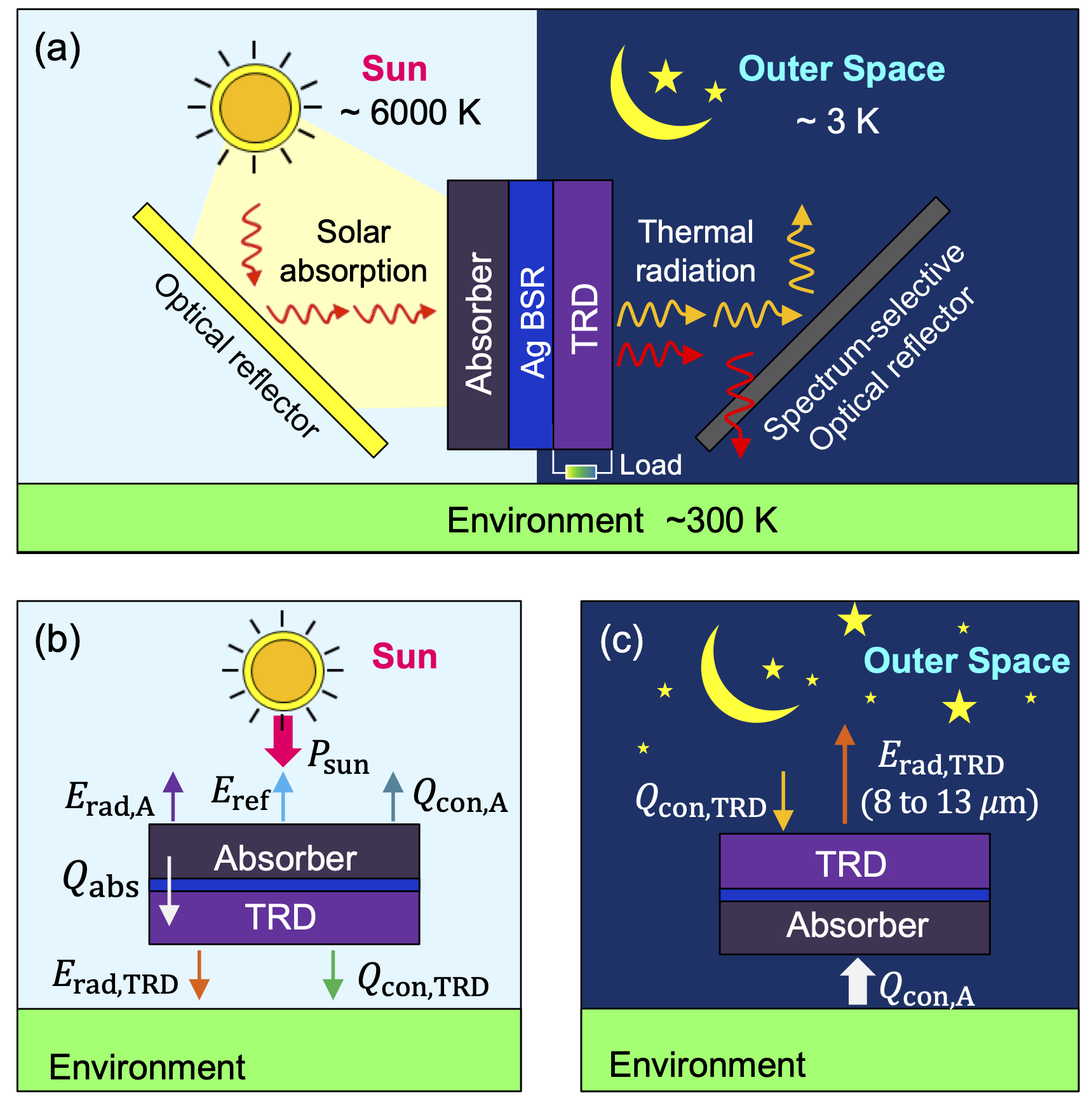}
\caption{(a) Schematic of a 24-hour TRD system for continuous electricity generation by optically coupling with the Sun, outer space, and environment. (b) and (c) correspond the energy flow diagram of daytime and nighttime TRD system.}
\label{fig:graph}
\end{figure}

A self-consistent thermodynamic model is developed for our proposed all-day TRD system. For simplicity, we use the subscripts “A” and “E” to denote absorber and environment, respectively. At daytime, the solar-to-electric conversion efficiency of the system is defined as
\begin{equation}
\eta=\frac{P}{P_{\text{sun}}}=\frac{JV}{C\int_{0}^{\infty} \mathcal{I}_{\text{AM1.5D}}(\lambda)\, d\lambda},
\end{equation}
where $P=JV$ represents the output electric power density, $J$ is the current density, and $V$ is the output voltage. 
$P_{\text{sun}}$ is the concentrated solar power flux impinging on the top surface of the absorber. $\mathcal{I}_{\text{AM1.5D}}$ represents the AM1.5D solar irradiance varying with the wavelength $\lambda$. $C$ denotes the solar concentration factor. 
The overall efficiency is also the product of the optothermal efficiency of the absorber ($\eta_{\text{ot}}$) and the thermoelectric conversion efficiency of the TRD ($\eta_{\text{TRD}}$), i.e., $\eta=\eta_{\text{ot}}\cdot \eta_{\text{TRD}}$. 
Here $\eta_{\text{ot}}=Q_{\text{abs}}/P_{\text{sun}}$ characterizes the solar-to-thermal conversion capacity of the absorber that converts the concentrated solar power flux into heat flux that flows into the TRD ($Q_{\text{abs}}$) and $\eta_{\text{TRD}}$ is defined as the ratio of the output electric power density to the incoming heat flux, i.e., $\eta_{\text{TRD}}=P/Q_{\text{abs}}$. To maximize the solar-to-thermal conversion efficiency, it is crucial to increase the total solar spectrum absorbance, while maintaining low thermal radiation loss. 

In the TRD, considering the effect of non-radiative (NR) process, the current density is given by \cite{strandberg2015theoretical,fernandez2018endoreversible,zhang2019design,fernandez2019theoretical}
\begin{equation}
J=e[N_{\text{abs}}(0,T_{\text{E}})-N_{\text{em}}(\Delta\mu,T_{\text{TRD}})]-J_{\text{NR}}(\Delta\mu,T_{\text{TRD}}),
\end{equation}
where $N(\Delta\mu,T)=\frac{c}{4\pi}\int_{0}^{\lambda_{\text{g}}} \epsilon_{\text{TRD}}(\lambda)D(\lambda)\Theta(\Delta\mu,T)\, d\lambda$ is the photon flux, $e$ is the electron charge, $c$ is the speed of light in vacuum, $T$ is the temperature, $\lambda_{\text{g}}$ is the cut-off wavelength corresponding to the semiconductor bandgap $E_{\text{g}}$, $\epsilon_{\text{TRD}}(\lambda)$ is the emissivity of the TRD or HgCdTe surface and, according to Kirchhoff’s law, is equal to the absorptivity, and $D(\lambda)=1/(c\pi^2\lambda^2)$ is the photon density of states in a homogeneous bulk material. 
$\Theta(\Delta\mu,T)=1/[e^{(\hbar c\lambda-\Delta\mu)/k_{\text{B}}T}-1]$ is the modified Bose-Einstein distributions of photons, and $\Delta\mu=e V$ is the chemical potential driving photon emission. 
The first term in Eq. (2) is related to the above-bandgap photon flux. 
The corresponding net radiative energy flux from the TRD to the environment is $E_{\text{rad, TRD}}=E_{\text{em}}(\Delta\mu,T_{\text{TRD}})-E_{\text{abs}}(0,T_{\text{E}})$, where $E(\Delta\mu,T)=\frac{\hbar c^2}{4\pi}\int_{0}^{\lambda_{\text{g}}} \frac{\epsilon_{\text{TRD}}(\lambda)D(\lambda)\Theta(\Delta\mu,T)}{\lambda}\, d\lambda$.

Furthermore, the non-radiative process, which serves as an additional loss mechanism, shall affect the net free carrier generation in the TRD. To quantitatively describe the non-radiative loss, including the Auger and Shockley-Read-Hall (SRH) process, the net current losses generated in the high-injection regime can be calculated as \cite{burger2018thin},
\begin{equation}
J_{\text{NR}}(\Delta\mu,T)=et\left[\tau_{\text{auger}}n_i^3\exp \left(\frac{3\Delta\mu}{2k_{\text{B}}T}\right)+\frac{n_i^2}{\tau_{\text{SRH}}}\exp\left(\frac{\Delta\mu}{2k_{\text{B}}T}\right)\right],
\end{equation}
where $t$ denotes the thickness of the active region, $\tau_{\text{auger}}$ represents the intrinsic Auger recombination times, $n_i$ is intrinsic carrier concentration, and $\tau_{\text{SRH}}$ is the bulk Shockley-Read-Hall lifetime based on the experimental measurements \cite{rogalski2005hgcdte}. 
Here, the parameter choice and temperature dependence of $\tau_{\text{auger}}$ and $n_i$ are based on the model for HgCdTe \cite{hansen1983calculation}. 
Eq. (3) illustrates that the non-radiative recombination is a volumetric effect because the number of non-radiative recombination events per unit area is proportional to the device’s thickness. We choose a thickness $t=100$ nm \cite{zhang2020designing}, such that there is sufficient thickness for photon absorption while the layer remains thin enough to mitigate the non-radiative effect.

The conversion efficiency and current density can be calculated from Eqs. (1-3) once the TRD’s temperature is determined from the energy balance condition. 
Taking into account all the power fluxes involving the absorber and TRD [Fig. 1(b) and 1(c)], we obtained,
\begin{subequations}
\label{eq:whole}
\begin{equation}
P_{\text{sun}}-E_{\text{rad,A}}-E_{\text{ref}}-Q_{\text{con,A}}-Q_{\text{abs}}=0, \label{subeq:2}
\end{equation}
\begin{equation}
Q_{\text{abs}}-E_{\text{rad, TRD}}-Q_{\text{con,TRD}}-P=0.\label{subeq:1}
\end{equation}
\end{subequations}
where $E_{\text{rad,A}}=\frac{2\pi \hbar c^2}{\lambda^5\int_{0}^{\infty} \epsilon_{\text{A}}(\lambda)[\Theta(T_{\text{A}})-\Theta(T_{\text{E}})]\, d\lambda}$ is the blackbody thermal radiation flux from the absorber into the environment, $\epsilon_{\text{A}}(\lambda)$ is the emissivity of the absorber’s top surface, $E_{\text{ref}}=C\int_{0}^{\infty} r(\lambda)\mathcal{I}_{\text{AM1.5D}}(\lambda)\, d\lambda$ is the reflected radiation flux of the absorber, $r(\lambda)=1-\epsilon_{\text{A}}(\lambda)$ represents the spectral reflectance [see Fig. 2(a)], $Q_{\text{con,A}}=h_{\text{A}}(T_{\text{A}}-T_{\text{E}})$ ($Q_{\text{con,TRD}}=h_{\text{TRD}}(T_{\text{TRD}}-T_{\text{E}})$) is the heat flux caused by heat convection and conduction between the absorber (TRD) and the environment according to the Newton heat transfer law, and $h_{\text{TRD}}=h_{\text{A}}=7$ W m$^{-2}$ K$^{-1}$ is the corresponding global heat transfer coefficient that is compatible to the standard air conditions.

At nighttime, the TRD system enables electric power generation by couple the cold outer space with the warm ambient environment through the atmospheric transparency window. 
The current density is given by
\begin{equation}
J=e[\Gamma(\lambda)F_{\text{abs}}(0,T_{\text{sky}})-F_{\text{em}}(\Delta\mu,T_{\text{TRD}})]-J_{\text{NR}}(\Delta\mu,T_{\text{TRD}}),
\end{equation}
where $F(\Delta\mu,T)=\frac{c}{4\pi}\int_{8\mu \text{m}}^{13\mu \text{m}} \epsilon_{\text{TRD}}(\lambda)D(\lambda)\Theta(\Delta\mu,T)\, d\lambda$, $T_{\text{sky}}$ is the temperature of the night sky, $\Gamma(\lambda)$ is the atmospheric transmittance in the zenith direction as obtained from MODTRAN5 \cite{berk2006modtran5}. 
The heat energy is extracted from environment or ambient air to the top surface of the absorber mainly by convection and conduction and is dissipated by heat radiative transfer, convection and conduction at the TRD surface. 
The power conversion efficiency can be expressed as
\begin{equation}
\eta=\frac{JV}{JV+E_{\text{rad,TRD}}(8-13\mu \text{m})+Q_{\text{con,TRD}}},
\end{equation}
where $E_{\text{rad,TRD}}(8-13\mu \text{m})$ is the net radiative energy flux with wavelength range of 8-13 $\mu$m. 
The temperature of the TRD at nighttime operation is variable, and is determined by the energy balance condition:
\begin{equation}
Q_{\text{con,A}}+JV+E_{\text{rad,TRD}}(8-13 \mu \text{m})+Q_{\text{con,TRD}}=0.
\end{equation}
We also consider the temperature-dependent bandgap of Hg$_\text{1-x}$Cd$_\text{x}$Te compounds, which can be described as $E_{\text{g}}(x,T)=-0.312+1.93x-0.81x^2+0.832x^3+5.35\times10^{-4}(1-2x)T$. We select a mole fraction of $x=0.12$ for Hg$_\text{1-x}$Cd$_\text{x}$Te, which corresponds to a low bandgap of 0.05 eV to 0.08 eV at the temperature range of 300-800 K, which enables significant above-bandgap emissivity in the atmospheric transparency window (0.09-0.15 eV). 


Figure 2 shows the optical properties of the components in the all-day TRD system, such as normalized AM 1.5 Direct solar spectrum, 3DSGM, the mid-infrared atmospheric transparency window, and HgCdTe. The nearly unity absorptivity ($r_{\text{A}}=0.96$) in the solar irradiance region (0.3 $\mu$m to 1.6 $\mu$m) ensures an excellent absorption of sunlight to heat up the TRD module under the Sun. The low reflectance allows the efficient absorbtion of most of the concentrated sunlight. Additionally, an excellent emissivity $\epsilon_{\text{TRD}}$ of HgCdTe (around 0.9) in the atmospheric transparency window ensures thermal infrared radiation with outer space. 
\begin{figure}[t]
\centering
\includegraphics[height=7.5cm]{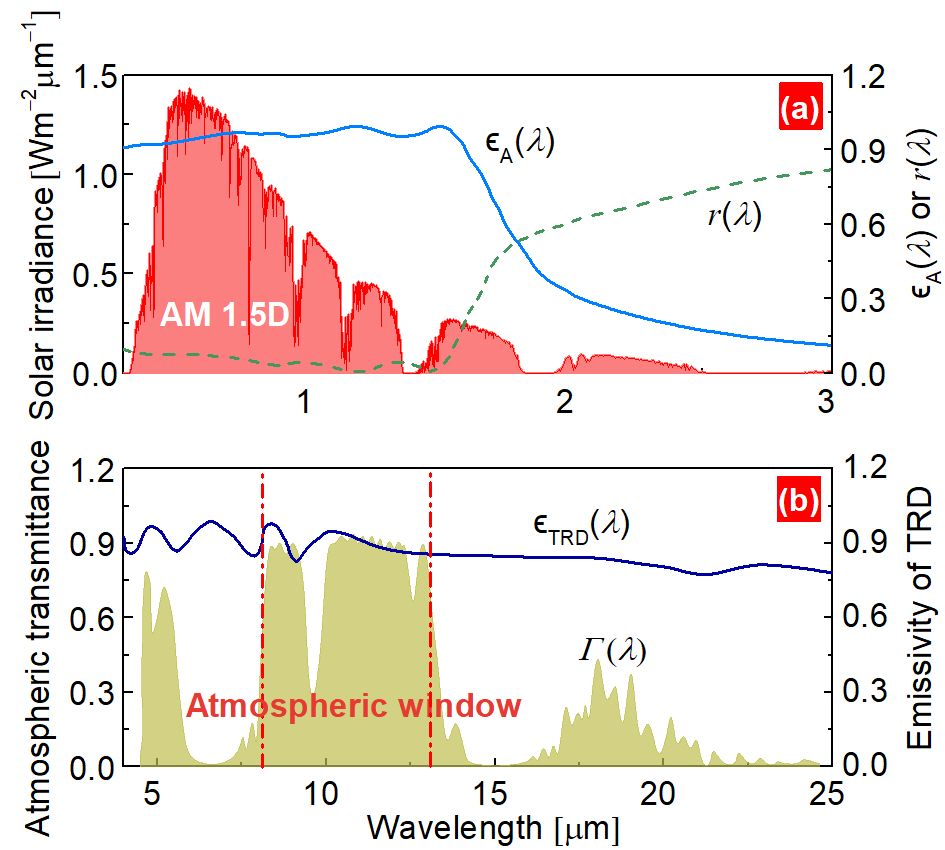}
\caption{Optical properties of the components in the system. (a) The solar irradiance of AM1.5 Direct solar spectrum, spectral emissivity $\epsilon_{\text{A}}(\lambda)$ and reflectance $r_{\text{A}}(\lambda)$ of the 3DSGM absorber; (b) atmospheric transmittance $\Gamma(\lambda)$ and emissivity of the TRD $\epsilon_{\text{TRD}}(\lambda)$ varying with the wavelength, where wavelength range from 8 to 13 $\mu$m corresponds to atmospheric window.}
\label{fig:graph}
\end{figure}

\begin{figure*}[t]
\centering
\includegraphics[height=11cm]{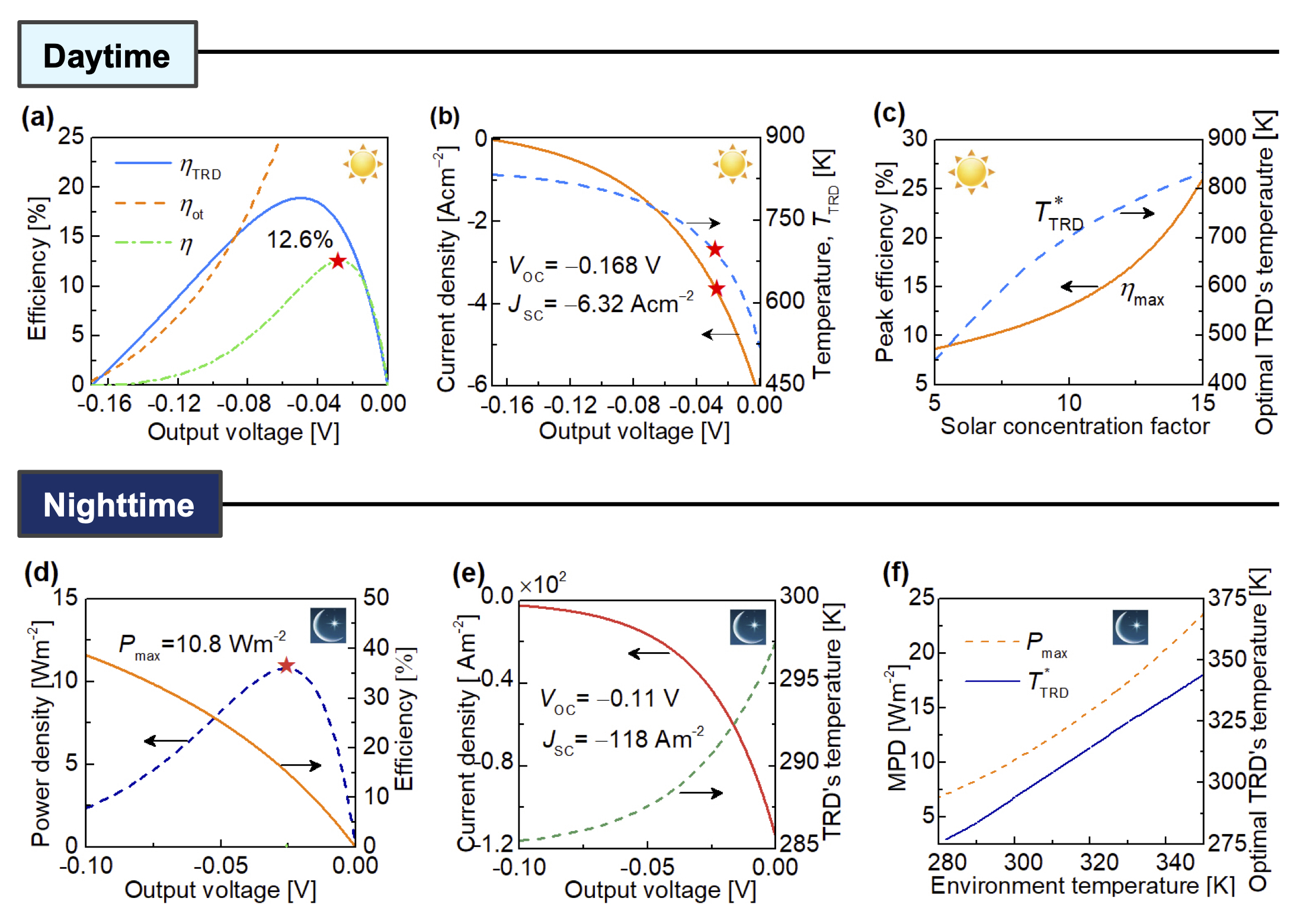}
\caption{The performance characteristics of the daytime TRD system under 10 suns (a-c) and nighttime TRD system operating between the environment at 300 K and the deep space at 3 K (e-f). Daytime: (a) The opto-thermal efficiency $\eta_{\text{ot}}$, TRD’s efficiency $\eta_{\text{TRD}}$, and overall efficiency $\eta$, (b) current density $J$ and TRD’s temperature $T_{\text{TRD}}$ as functions of the negative output voltage $V$, and (c) peak efficiency (PE) and corresponding optimum temperature under different solar concentration factors. Nighttime: (d) The power density $P$ and conversion efficiency $\eta$, (e) current density $J$ and TRD’s temperature $T_{\text{TRD}}$ as functions of the negative output voltage $V$, and (f) maximum power density (MPD) and corresponding TRD’s temperature $T^{*}_{\text{TRD}}$ under different environment temperatures.}
\label{fig:graph}
\end{figure*}

Using Eqs. (1-4), we simulate realistically the performance characteristics of the daytime TRD system [Fig. 3(a)-3(b)]. 
Fig. 3(a) shows the efficiency of the daytime TRD system and subsystems as a function of the output voltage. 
Individually, the opto-thermal efficiency and the TRD’s efficiency reaches 92.8\% and 18.5\%, respectively, thus revealing the TRD as the bottleneck that limits the conversion efficiency during daytime operation. 
The product of these two individual efficiencies, i.e., 78.2\% and 16.1\%, yields the combined peak efficiency (PE) of 12.6\% under 10 suns. 
Fig. 3(b) obtains the volt-ampere characteristic curve of the daytime TRD system. The current density decreases with an increasing output voltage, and the system exhibits an open-circuit voltage of $-0.168$ V and a short-circuit current density of $-6.32$ Acm$^{-2}$. Here the operating current density and voltages have opposite signs in contrast to photovoltaic cell because the radiative processes are reciprocal. Since the energy exchange of the TRD is closely dependent on the net radiative energy fluxes from the TRD, the TRD’s temperature decreases at higher output voltages [Fig. 3(b)] to satisfy the energy balance in Eq. 4, where the radiative energy fluxes transported from the TRD to the environment is substantially lower.

We further identify the process that contributes to the major energy losses in the TRD. 
We calculate the PEs of the system in four cases, including: (1) the ideal system, (2) the system with only the optical loss, (3) the system with only the non-radiative loss, (4) the system with only the heat conduction and convection loss. The corresponding values are 28.6\%, 25.4\%, 22.9\%, 27.3\%. 
Compared with the proposed system with all losses (12.6\%), we identify the non-radiative recombination is a main loss mechanism that significantly degrades the daytime TRD system performance. 
The ongoing research efforts suppress the Auger process using quantum wells, doping profile control, superlattices, or heterostructures and, meanwhile, mitigate surface recombination by passivation. 
Moreover, to reduce the optical loss, a back-reflector metallic grating (Ag mirror) can be used to promote light utilization in the active layer, which is thermally conducting on the bottom surface. 
A device module encapsulation can also be used to protect the system from the environment and to restrict conductive and convective heat loss.

Next, we study how different solar concentration factors can affect the optimal performance characteristics of the daytime TRD system, as shown in Fig. 3(c). 
In general, larger solar concentrations correspond to more input energy fluxes, and thus lead to higher PEs ($\eta_{\text{max}}$) and corresponding TRD's temperatures ($T^{*}_{\text{TRD}}$). 
Further increment in overall efficiency is possible with higher optical concentration for the daytime TRD system to operate at higher temperatures.
Finally, we compare the PEs and the corresponding working temperatures between the daytime TRD system and the concentrated TEG-based solar cell, in which a PE of 6\% and 520 K has been theoretically achieved \cite{chen2011theoretical}. 
The theoretically predicted values of the daytime TRD system, i.e., 8\% and 520 K when irreversible losses are included, is higher than TEG-based solar cells. 
In addition, 20\% at 800 K of the low-concentration (13 suns) TRD system are substantially higher than those of the high-concentration solar cells, such as solar thermophotovoltaics (8.4\% at 1676 K) \cite{bhatt2020high} and solar thermionics (7\% at 2000 K) \cite{rouklove1966thermionic}. 
These performance figure-of-merits demonstrate the potential advantageous of daytime TRD system in solar-to-electric energy conversion.

The operation of TRD at nighttime is different from that at daytime. In the latter, the state of the maximum efficiency coincides with that of the maximum power density (MPD) because the efficiency is defined as the ratio of the output electric power to the total incident power per unit area, which is a given quantity ($\sim1003$ Wm$^{-2}$). 
For nighttime TRD operation, the absorbed heat flux is variable and is closely dependent on the parameters of the TRD, such as the voltage and bandgap. The power density exhibits a non-monotonous dependence with the output voltage [Fig. 3(d)]. In contrast, the conversion efficiency decreases monotonously with the output voltage because the input heat flux or TRD’s temperature increases with the output voltage [Fig. 3(e)]. The numerical results reveal that the nighttime TRD system can generate an MPD of 10.8 Wm$^{-2}$ for a voltage of $-0.026$ V, while yielding a corresponding efficiency of 5\%. Fig. 3(e) shows the volt-ampere characteristic curve of the nighttime TRD system. The current density increases with the increment of the output voltage, and the system exhibits an open-circuit voltage of $-0.11$ V and a short-circuit current of $-118$ Am$^{-2}$. 

We further note that the MPD and the corresponding optimal values of the TRD temperature are depend on the temperature of the ambient environment. 
Taking MPD as the key objective function, we find that the MPD and the TRD temperature monotonically increases with increasing temperature in the range of 280 K to 350 K [Fig. 3(f)]. 
Comparing the nighttime TRD system with the nighttime TEG radiative cooling system based on a commercial TEG module (TG12–4-01LS, Marlow Industries) \cite{raman2019generating}, the MPD of the nighttime TRD system proposed here is 10.8 Wm$^{-2}$, which is substantially higher than the 52 mWm$^{-2}$ \cite{fan2020maximal} and 20 mWm$^{-2}$ \cite{zhao2020modeling} at room temperature of the nighttime TEG. 
This comparison suggest the superior nighttime energy harvesting performance of the TRD system proposed. 

In summary, we proposed a high-performance 24-hour TRD system enabled by HgCdTe p-n junctions for 24-hour harvesting energy from the sun and outer space
The system yields a PE of 12.6\% under low concentration factors at daytime and a maximum power density of 10.8 Wm$^{-2}$ at nighttime room temperature, which significantly outperformed other representative TEG-based energy harvesting devices. 
The device model developed above and the device performance characteristics outlined in Fig. 3 shall offer an important design guideline for developing high-performance TRD-based energy converters. 
Our results reveal an alternative TRD-based route towards efficient 24-hour electricity generation.

This work is supported by the the National Natural Science Foundation of China (61974056), the Key Research and Development Program of Jiangsu Province (BE2020756), and Ministry of Education-Singapore (2018-T2-1-007).

\section*{data availability}
The data that support the findings of this study are available from the corresponding author upon reasonable request.

\nocite{*}
\section*{References}
\bibliography{sample}

\begin{thebibliography}{36}%
\makeatletter
\providecommand \@ifxundefined [1]{%
 \@ifx{#1\undefined}
}%
\providecommand \@ifnum [1]{%
 \ifnum #1\expandafter \@firstoftwo
 \else \expandafter \@secondoftwo
 \fi
}%
\providecommand \@ifx [1]{%
 \ifx #1\expandafter \@firstoftwo
 \else \expandafter \@secondoftwo
 \fi
}%
\providecommand \natexlab [1]{#1}%
\providecommand \enquote  [1]{``#1''}%
\providecommand \bibnamefont  [1]{#1}%
\providecommand \bibfnamefont [1]{#1}%
\providecommand \citenamefont [1]{#1}%
\providecommand \href@noop [0]{\@secondoftwo}%
\providecommand \href [0]{\begingroup \@sanitize@url \@href}%
\providecommand \@href[1]{\@@startlink{#1}\@@href}%
\providecommand \@@href[1]{\endgroup#1\@@endlink}%
\providecommand \@sanitize@url [0]{\catcode `\\12\catcode `\$12\catcode
  `\&12\catcode `\#12\catcode `\^12\catcode `\_12\catcode `\%12\relax}%
\providecommand \@@startlink[1]{}%
\providecommand \@@endlink[0]{}%
\providecommand \url  [0]{\begingroup\@sanitize@url \@url }%
\providecommand \@url [1]{\endgroup\@href {#1}{\urlprefix }}%
\providecommand \urlprefix  [0]{URL }%
\providecommand \Eprint [0]{\href }%
\providecommand \doibase [0]{http://dx.doi.org/}%
\providecommand \selectlanguage [0]{\@gobble}%
\providecommand \bibinfo  [0]{\@secondoftwo}%
\providecommand \bibfield  [0]{\@secondoftwo}%
\providecommand \translation [1]{[#1]}%
\providecommand \BibitemOpen [0]{}%
\providecommand \bibitemStop [0]{}%
\providecommand \bibitemNoStop [0]{.\EOS\space}%
\providecommand \EOS [0]{\spacefactor3000\relax}%
\providecommand \BibitemShut  [1]{\csname bibitem#1\endcsname}%
\let\auto@bib@innerbib\@empty
\bibitem [{\citenamefont {Chu}, \citenamefont {Cui},\ and\ \citenamefont
  {Liu}(2017)}]{chu2017path}%
  \BibitemOpen
  \bibfield  {author} {\bibinfo {author} {\bibfnamefont {S.}~\bibnamefont
  {Chu}}, \bibinfo {author} {\bibfnamefont {Y.}~\bibnamefont {Cui}}, \ and\
  \bibinfo {author} {\bibfnamefont {N.}~\bibnamefont {Liu}},\ }\href@noop {}
  {\bibfield  {journal} {\bibinfo  {journal} {Nature Materials}\ }\textbf
  {\bibinfo {volume} {16}},\ \bibinfo {pages} {16} (\bibinfo {year}
  {2017})}\BibitemShut {NoStop}%
\bibitem [{\citenamefont {Shindell}\ and\ \citenamefont
  {Smith}(2019)}]{shindell2019climate}%
  \BibitemOpen
  \bibfield  {author} {\bibinfo {author} {\bibfnamefont {D.}~\bibnamefont
  {Shindell}}\ and\ \bibinfo {author} {\bibfnamefont {C.~J.}\ \bibnamefont
  {Smith}},\ }\href@noop {} {\bibfield  {journal} {\bibinfo  {journal}
  {Nature}\ }\textbf {\bibinfo {volume} {573}},\ \bibinfo {pages} {408}
  (\bibinfo {year} {2019})}\BibitemShut {NoStop}%
\bibitem [{\citenamefont {Bagdadee}\ and\ \citenamefont
  {Zhang}(2020)}]{bagdadee2020electrical}%
  \BibitemOpen
  \bibfield  {author} {\bibinfo {author} {\bibfnamefont {A.~H.}\ \bibnamefont
  {Bagdadee}}\ and\ \bibinfo {author} {\bibfnamefont {L.}~\bibnamefont
  {Zhang}},\ }\href@noop {} {\bibfield  {journal} {\bibinfo  {journal} {Energy
  Reports}\ }\textbf {\bibinfo {volume} {6}},\ \bibinfo {pages} {480} (\bibinfo
  {year} {2020})}\BibitemShut {NoStop}%
\bibitem [{\citenamefont {Wu}\ \emph {et~al.}(2020)\citenamefont {Wu},
  \citenamefont {Wang}, \citenamefont {Chen}, \citenamefont {Wu}, \citenamefont
  {Shen}, \citenamefont {Lin}, \citenamefont {Ge}, \citenamefont {Hu},
  \citenamefont {Zhang}, \citenamefont {Meng} \emph {et~al.}}]{wu2020two}%
  \BibitemOpen
  \bibfield  {author} {\bibinfo {author} {\bibfnamefont {G.}~\bibnamefont
  {Wu}}, \bibinfo {author} {\bibfnamefont {X.}~\bibnamefont {Wang}}, \bibinfo
  {author} {\bibfnamefont {Y.}~\bibnamefont {Chen}}, \bibinfo {author}
  {\bibfnamefont {S.}~\bibnamefont {Wu}}, \bibinfo {author} {\bibfnamefont
  {H.}~\bibnamefont {Shen}}, \bibinfo {author} {\bibfnamefont {T.}~\bibnamefont
  {Lin}}, \bibinfo {author} {\bibfnamefont {J.}~\bibnamefont {Ge}}, \bibinfo
  {author} {\bibfnamefont {W.}~\bibnamefont {Hu}}, \bibinfo {author}
  {\bibfnamefont {S.-T.}\ \bibnamefont {Zhang}}, \bibinfo {author}
  {\bibfnamefont {X.}~\bibnamefont {Meng}},  \emph {et~al.},\ }\href@noop {}
  {\bibfield  {journal} {\bibinfo  {journal} {Applied Physics Letters}\
  }\textbf {\bibinfo {volume} {116}},\ \bibinfo {pages} {073101} (\bibinfo
  {year} {2020})}\BibitemShut {NoStop}%
\bibitem [{\citenamefont {Qu}, \citenamefont {Hong},\ and\ \citenamefont
  {Jin}(2019)}]{qu2019spectral}%
  \BibitemOpen
  \bibfield  {author} {\bibinfo {author} {\bibfnamefont {W.}~\bibnamefont
  {Qu}}, \bibinfo {author} {\bibfnamefont {H.}~\bibnamefont {Hong}}, \ and\
  \bibinfo {author} {\bibfnamefont {H.}~\bibnamefont {Jin}},\ }\href@noop {}
  {\bibfield  {journal} {\bibinfo  {journal} {Applied Energy}\ }\textbf
  {\bibinfo {volume} {248}},\ \bibinfo {pages} {162} (\bibinfo {year}
  {2019})}\BibitemShut {NoStop}%
\bibitem [{\citenamefont {Hoppmann}\ \emph {et~al.}(2014)\citenamefont
  {Hoppmann}, \citenamefont {Volland}, \citenamefont {Schmidt},\ and\
  \citenamefont {Hoffmann}}]{hoppmann2014economic}%
  \BibitemOpen
  \bibfield  {author} {\bibinfo {author} {\bibfnamefont {J.}~\bibnamefont
  {Hoppmann}}, \bibinfo {author} {\bibfnamefont {J.}~\bibnamefont {Volland}},
  \bibinfo {author} {\bibfnamefont {T.~S.}\ \bibnamefont {Schmidt}}, \ and\
  \bibinfo {author} {\bibfnamefont {V.~H.}\ \bibnamefont {Hoffmann}},\
  }\href@noop {} {\bibfield  {journal} {\bibinfo  {journal} {Renewable and
  Sustainable Energy Reviews}\ }\textbf {\bibinfo {volume} {39}},\ \bibinfo
  {pages} {1101} (\bibinfo {year} {2014})}\BibitemShut {NoStop}%
\bibitem [{\citenamefont {Xu}, \citenamefont {Li},\ and\ \citenamefont
  {Chan}(2015)}]{xu2015application}%
  \BibitemOpen
  \bibfield  {author} {\bibinfo {author} {\bibfnamefont {B.}~\bibnamefont
  {Xu}}, \bibinfo {author} {\bibfnamefont {P.}~\bibnamefont {Li}}, \ and\
  \bibinfo {author} {\bibfnamefont {C.}~\bibnamefont {Chan}},\ }\href@noop {}
  {\bibfield  {journal} {\bibinfo  {journal} {Applied Energy}\ }\textbf
  {\bibinfo {volume} {160}},\ \bibinfo {pages} {286} (\bibinfo {year}
  {2015})}\BibitemShut {NoStop}%
\bibitem [{\citenamefont {Cabraal}, \citenamefont {Barnes},\ and\ \citenamefont
  {Agarwal}(2005)}]{cabraal2005productive}%
  \BibitemOpen
  \bibfield  {author} {\bibinfo {author} {\bibfnamefont {R.~A.}\ \bibnamefont
  {Cabraal}}, \bibinfo {author} {\bibfnamefont {D.~F.}\ \bibnamefont {Barnes}},
  \ and\ \bibinfo {author} {\bibfnamefont {S.~G.}\ \bibnamefont {Agarwal}},\
  }\href@noop {} {\bibfield  {journal} {\bibinfo  {journal} {Annu. Rev.
  Environ. Resour.}\ }\textbf {\bibinfo {volume} {30}},\ \bibinfo {pages} {117}
  (\bibinfo {year} {2005})}\BibitemShut {NoStop}%
\bibitem [{\citenamefont {Yin}\ \emph {et~al.}(2020)\citenamefont {Yin},
  \citenamefont {Yang}, \citenamefont {Tan},\ and\ \citenamefont
  {Fan}}]{yin2020terrestrial}%
  \BibitemOpen
  \bibfield  {author} {\bibinfo {author} {\bibfnamefont {X.}~\bibnamefont
  {Yin}}, \bibinfo {author} {\bibfnamefont {R.}~\bibnamefont {Yang}}, \bibinfo
  {author} {\bibfnamefont {G.}~\bibnamefont {Tan}}, \ and\ \bibinfo {author}
  {\bibfnamefont {S.}~\bibnamefont {Fan}},\ }\href@noop {} {\bibfield
  {journal} {\bibinfo  {journal} {Science}\ }\textbf {\bibinfo {volume}
  {370}},\ \bibinfo {pages} {786} (\bibinfo {year} {2020})}\BibitemShut
  {NoStop}%
\bibitem [{\citenamefont {Liu}\ \emph {et~al.}(2020)\citenamefont {Liu},
  \citenamefont {Zhang}, \citenamefont {Tang}, \citenamefont {Zhou},
  \citenamefont {Zhang}, \citenamefont {Ye},\ and\ \citenamefont
  {Zhao}}]{liu2020recent}%
  \BibitemOpen
  \bibfield  {author} {\bibinfo {author} {\bibfnamefont {J.}~\bibnamefont
  {Liu}}, \bibinfo {author} {\bibfnamefont {J.}~\bibnamefont {Zhang}}, \bibinfo
  {author} {\bibfnamefont {H.}~\bibnamefont {Tang}}, \bibinfo {author}
  {\bibfnamefont {Z.}~\bibnamefont {Zhou}}, \bibinfo {author} {\bibfnamefont
  {D.}~\bibnamefont {Zhang}}, \bibinfo {author} {\bibfnamefont
  {L.}~\bibnamefont {Ye}}, \ and\ \bibinfo {author} {\bibfnamefont
  {D.}~\bibnamefont {Zhao}},\ }\href@noop {} {\bibfield  {journal} {\bibinfo
  {journal} {Nano Energy}\ }\textbf {\bibinfo {volume} {81}},\ \bibinfo {pages}
  {105611} (\bibinfo {year} {2020})}\BibitemShut {NoStop}%
\bibitem [{\citenamefont {Raman}, \citenamefont {Li},\ and\ \citenamefont
  {Fan}(2019)}]{raman2019generating}%
  \BibitemOpen
  \bibfield  {author} {\bibinfo {author} {\bibfnamefont {A.~P.}\ \bibnamefont
  {Raman}}, \bibinfo {author} {\bibfnamefont {W.}~\bibnamefont {Li}}, \ and\
  \bibinfo {author} {\bibfnamefont {S.}~\bibnamefont {Fan}},\ }\href@noop {}
  {\bibfield  {journal} {\bibinfo  {journal} {Joule}\ }\textbf {\bibinfo
  {volume} {3}},\ \bibinfo {pages} {2679} (\bibinfo {year} {2019})}\BibitemShut
  {NoStop}%
\bibitem [{\citenamefont {Fan}\ \emph {et~al.}(2020)\citenamefont {Fan},
  \citenamefont {Li}, \citenamefont {Jin}, \citenamefont {Orenstein},\ and\
  \citenamefont {Fan}}]{fan2020maximal}%
  \BibitemOpen
  \bibfield  {author} {\bibinfo {author} {\bibfnamefont {L.}~\bibnamefont
  {Fan}}, \bibinfo {author} {\bibfnamefont {W.}~\bibnamefont {Li}}, \bibinfo
  {author} {\bibfnamefont {W.}~\bibnamefont {Jin}}, \bibinfo {author}
  {\bibfnamefont {M.}~\bibnamefont {Orenstein}}, \ and\ \bibinfo {author}
  {\bibfnamefont {S.}~\bibnamefont {Fan}},\ }\href@noop {} {\bibfield
  {journal} {\bibinfo  {journal} {Optics Express}\ }\textbf {\bibinfo {volume}
  {28}},\ \bibinfo {pages} {25460} (\bibinfo {year} {2020})}\BibitemShut
  {NoStop}%
\bibitem [{\citenamefont {Zhao}, \citenamefont {Pei},\ and\ \citenamefont
  {Raman}(2020)}]{zhao2020modeling}%
  \BibitemOpen
  \bibfield  {author} {\bibinfo {author} {\bibfnamefont {B.}~\bibnamefont
  {Zhao}}, \bibinfo {author} {\bibfnamefont {G.}~\bibnamefont {Pei}}, \ and\
  \bibinfo {author} {\bibfnamefont {A.~P.}\ \bibnamefont {Raman}},\ }\href@noop
  {} {\bibfield  {journal} {\bibinfo  {journal} {Applied Physics Letters}\
  }\textbf {\bibinfo {volume} {117}},\ \bibinfo {pages} {163903} (\bibinfo
  {year} {2020})}\BibitemShut {NoStop}%
\bibitem [{\citenamefont {Chen}\ \emph {et~al.}(2019)\citenamefont {Chen},
  \citenamefont {Zhu}, \citenamefont {Li},\ and\ \citenamefont
  {Fan}}]{chen2019simultaneously}%
  \BibitemOpen
  \bibfield  {author} {\bibinfo {author} {\bibfnamefont {Z.}~\bibnamefont
  {Chen}}, \bibinfo {author} {\bibfnamefont {L.}~\bibnamefont {Zhu}}, \bibinfo
  {author} {\bibfnamefont {W.}~\bibnamefont {Li}}, \ and\ \bibinfo {author}
  {\bibfnamefont {S.}~\bibnamefont {Fan}},\ }\href@noop {} {\bibfield
  {journal} {\bibinfo  {journal} {Joule}\ }\textbf {\bibinfo {volume} {3}},\
  \bibinfo {pages} {101} (\bibinfo {year} {2019})}\BibitemShut {NoStop}%
\bibitem [{\citenamefont {Tian}\ \emph {et~al.}(2020)\citenamefont {Tian},
  \citenamefont {Liu}, \citenamefont {Chen},\ and\ \citenamefont
  {Zheng}}]{tian2020harvesting}%
  \BibitemOpen
  \bibfield  {author} {\bibinfo {author} {\bibfnamefont {Y.}~\bibnamefont
  {Tian}}, \bibinfo {author} {\bibfnamefont {X.}~\bibnamefont {Liu}}, \bibinfo
  {author} {\bibfnamefont {F.}~\bibnamefont {Chen}}, \ and\ \bibinfo {author}
  {\bibfnamefont {Y.}~\bibnamefont {Zheng}},\ }\href@noop {} {\bibfield
  {journal} {\bibinfo  {journal} {Scientific Reports}\ }\textbf {\bibinfo
  {volume} {10}},\ \bibinfo {pages} {1} (\bibinfo {year} {2020})}\BibitemShut
  {NoStop}%
\bibitem [{\citenamefont {Ao}\ \emph {et~al.}(2020)\citenamefont {Ao},
  \citenamefont {Li}, \citenamefont {Zhao}, \citenamefont {Hu}, \citenamefont
  {Ren}, \citenamefont {Yang}, \citenamefont {Liu}, \citenamefont {Cao},
  \citenamefont {Feng}, \citenamefont {Yang} \emph {et~al.}}]{ao2020spectral}%
  \BibitemOpen
  \bibfield  {author} {\bibinfo {author} {\bibfnamefont {X.}~\bibnamefont
  {Ao}}, \bibinfo {author} {\bibfnamefont {B.}~\bibnamefont {Li}}, \bibinfo
  {author} {\bibfnamefont {B.}~\bibnamefont {Zhao}}, \bibinfo {author}
  {\bibfnamefont {M.}~\bibnamefont {Hu}}, \bibinfo {author} {\bibfnamefont
  {H.}~\bibnamefont {Ren}}, \bibinfo {author} {\bibfnamefont {H.}~\bibnamefont
  {Yang}}, \bibinfo {author} {\bibfnamefont {J.}~\bibnamefont {Liu}}, \bibinfo
  {author} {\bibfnamefont {J.}~\bibnamefont {Cao}}, \bibinfo {author}
  {\bibfnamefont {J.}~\bibnamefont {Feng}}, \bibinfo {author} {\bibfnamefont
  {Y.}~\bibnamefont {Yang}},  \emph {et~al.},\ }\href@noop {} {\bibfield
  {journal} {\bibinfo  {journal} {arXiv preprint arXiv:2004.00459}\ } (\bibinfo
  {year} {2020})}\BibitemShut {NoStop}%
\bibitem [{\citenamefont {Xia}\ \emph {et~al.}(2020)\citenamefont {Xia},
  \citenamefont {Zhang}, \citenamefont {Meng},\ and\ \citenamefont
  {Yu}}]{xia202024}%
  \BibitemOpen
  \bibfield  {author} {\bibinfo {author} {\bibfnamefont {Z.}~\bibnamefont
  {Xia}}, \bibinfo {author} {\bibfnamefont {Z.}~\bibnamefont {Zhang}}, \bibinfo
  {author} {\bibfnamefont {Z.}~\bibnamefont {Meng}}, \ and\ \bibinfo {author}
  {\bibfnamefont {Z.}~\bibnamefont {Yu}},\ }\href@noop {} {\bibfield  {journal}
  {\bibinfo  {journal} {Journal of Quantitative Spectroscopy and Radiative
  Transfer}\ }\textbf {\bibinfo {volume} {251}},\ \bibinfo {pages} {107038}
  (\bibinfo {year} {2020})}\BibitemShut {NoStop}%
\bibitem [{\citenamefont {Liu}\ \emph {et~al.}(2021)\citenamefont {Liu},
  \citenamefont {Zhang}, \citenamefont {Yuan}, \citenamefont {Zhang},
  \citenamefont {Xing},\ and\ \citenamefont {Zhou}}]{liu220model}%
  \BibitemOpen
  \bibfield  {author} {\bibinfo {author} {\bibfnamefont {J.}~\bibnamefont
  {Liu}}, \bibinfo {author} {\bibfnamefont {J.}~\bibnamefont {Zhang}}, \bibinfo
  {author} {\bibfnamefont {J.}~\bibnamefont {Yuan}}, \bibinfo {author}
  {\bibfnamefont {D.}~\bibnamefont {Zhang}}, \bibinfo {author} {\bibfnamefont
  {J.}~\bibnamefont {Xing}}, \ and\ \bibinfo {author} {\bibfnamefont
  {Z.}~\bibnamefont {Zhou}},\ }\href@noop {} {\bibfield  {journal} {\bibinfo
  {journal} {Solar Energy Materials and Solar Cells}\ }\textbf {\bibinfo
  {volume} {220}},\ \bibinfo {pages} {110855} (\bibinfo {year}
  {2021})}\BibitemShut {NoStop}%
\bibitem [{\citenamefont {Ishii}, \citenamefont {Dao},\ and\ \citenamefont
  {Nagao}(2020)}]{ishii2020radiative}%
  \BibitemOpen
  \bibfield  {author} {\bibinfo {author} {\bibfnamefont {S.}~\bibnamefont
  {Ishii}}, \bibinfo {author} {\bibfnamefont {T.~D.}\ \bibnamefont {Dao}}, \
  and\ \bibinfo {author} {\bibfnamefont {T.}~\bibnamefont {Nagao}},\
  }\href@noop {} {\bibfield  {journal} {\bibinfo  {journal} {Applied Physics
  Letters}\ }\textbf {\bibinfo {volume} {117}},\ \bibinfo {pages} {013901}
  (\bibinfo {year} {2020})}\BibitemShut {NoStop}%
\bibitem [{\citenamefont {Strandberg}(2015)}]{strandberg2015theoretical}%
  \BibitemOpen
  \bibfield  {author} {\bibinfo {author} {\bibfnamefont {R.}~\bibnamefont
  {Strandberg}},\ }\href@noop {} {\bibfield  {journal} {\bibinfo  {journal}
  {Journal of Applied Physics}\ }\textbf {\bibinfo {volume} {117}},\ \bibinfo
  {pages} {055105} (\bibinfo {year} {2015})}\BibitemShut {NoStop}%
\bibitem [{\citenamefont {Fern{\'a}ndez}(2018)}]{fernandez2018endoreversible}%
  \BibitemOpen
  \bibfield  {author} {\bibinfo {author} {\bibfnamefont {J.~J.}\ \bibnamefont
  {Fern{\'a}ndez}},\ }\href@noop {} {\bibfield  {journal} {\bibinfo  {journal}
  {Journal of Applied Physics}\ }\textbf {\bibinfo {volume} {123}},\ \bibinfo
  {pages} {164501} (\bibinfo {year} {2018})}\BibitemShut {NoStop}%
\bibitem [{\citenamefont {Zhang}\ \emph {et~al.}(2019)\citenamefont {Zhang},
  \citenamefont {Ang}, \citenamefont {can Chen},\ and\ \citenamefont
  {Ang}}]{zhang2019design}%
  \BibitemOpen
  \bibfield  {author} {\bibinfo {author} {\bibfnamefont {X.}~\bibnamefont
  {Zhang}}, \bibinfo {author} {\bibfnamefont {Y.~S.}\ \bibnamefont {Ang}},
  \bibinfo {author} {\bibfnamefont {J.}~\bibnamefont {can Chen}}, \ and\
  \bibinfo {author} {\bibfnamefont {L.~K.}\ \bibnamefont {Ang}},\ }\href@noop
  {} {\bibfield  {journal} {\bibinfo  {journal} {Optics Letters}\ }\textbf
  {\bibinfo {volume} {44}},\ \bibinfo {pages} {3354} (\bibinfo {year}
  {2019})}\BibitemShut {NoStop}%
\bibitem [{\citenamefont {Fern{\'a}ndez}(2019)}]{fernandez2019theoretical}%
  \BibitemOpen
  \bibfield  {author} {\bibinfo {author} {\bibfnamefont {J.}~\bibnamefont
  {Fern{\'a}ndez}},\ }\href@noop {} {\bibfield  {journal} {\bibinfo  {journal}
  {Journal of Applied Physics}\ }\textbf {\bibinfo {volume} {125}},\ \bibinfo
  {pages} {103101} (\bibinfo {year} {2019})}\BibitemShut {NoStop}%
\bibitem [{\citenamefont {Santhanam}\ and\ \citenamefont
  {Fan}(2016)}]{santhanam2016thermal}%
  \BibitemOpen
  \bibfield  {author} {\bibinfo {author} {\bibfnamefont {P.}~\bibnamefont
  {Santhanam}}\ and\ \bibinfo {author} {\bibfnamefont {S.}~\bibnamefont
  {Fan}},\ }\href@noop {} {\bibfield  {journal} {\bibinfo  {journal} {Physical
  Review B}\ }\textbf {\bibinfo {volume} {93}},\ \bibinfo {pages} {161410}
  (\bibinfo {year} {2016})}\BibitemShut {NoStop}%
\bibitem [{\citenamefont {Buddhiraju}, \citenamefont {Santhanam},\ and\
  \citenamefont {Fan}(2018)}]{buddhiraju2018thermodynamic}%
  \BibitemOpen
  \bibfield  {author} {\bibinfo {author} {\bibfnamefont {S.}~\bibnamefont
  {Buddhiraju}}, \bibinfo {author} {\bibfnamefont {P.}~\bibnamefont
  {Santhanam}}, \ and\ \bibinfo {author} {\bibfnamefont {S.}~\bibnamefont
  {Fan}},\ }\href@noop {} {\bibfield  {journal} {\bibinfo  {journal}
  {Proceedings of the National Academy of Sciences}\ }\textbf {\bibinfo
  {volume} {115}},\ \bibinfo {pages} {E3609} (\bibinfo {year}
  {2018})}\BibitemShut {NoStop}%
\bibitem [{\citenamefont {Deppe}\ and\ \citenamefont
  {Munday}(2019)}]{deppe2019nighttime}%
  \BibitemOpen
  \bibfield  {author} {\bibinfo {author} {\bibfnamefont {T.}~\bibnamefont
  {Deppe}}\ and\ \bibinfo {author} {\bibfnamefont {J.~N.}\ \bibnamefont
  {Munday}},\ }\href@noop {} {\bibfield  {journal} {\bibinfo  {journal} {ACS
  Photonics}\ }\textbf {\bibinfo {volume} {7}},\ \bibinfo {pages} {1} (\bibinfo
  {year} {2019})}\BibitemShut {NoStop}%
\bibitem [{\citenamefont {Li}, \citenamefont {Buddhiraju},\ and\ \citenamefont
  {Fan}(2020)}]{li2020thermodynamic}%
  \BibitemOpen
  \bibfield  {author} {\bibinfo {author} {\bibfnamefont {W.}~\bibnamefont
  {Li}}, \bibinfo {author} {\bibfnamefont {S.}~\bibnamefont {Buddhiraju}}, \
  and\ \bibinfo {author} {\bibfnamefont {S.}~\bibnamefont {Fan}},\ }\href@noop
  {} {\bibfield  {journal} {\bibinfo  {journal} {Light: Science \&
  Applications}\ }\textbf {\bibinfo {volume} {9}},\ \bibinfo {pages} {1}
  (\bibinfo {year} {2020})}\BibitemShut {NoStop}%
\bibitem [{\citenamefont {Lin}\ \emph {et~al.}(2020)\citenamefont {Lin},
  \citenamefont {Lin}, \citenamefont {Yang},\ and\ \citenamefont
  {Jia}}]{lin2020structured}%
  \BibitemOpen
  \bibfield  {author} {\bibinfo {author} {\bibfnamefont {K.-T.}\ \bibnamefont
  {Lin}}, \bibinfo {author} {\bibfnamefont {H.}~\bibnamefont {Lin}}, \bibinfo
  {author} {\bibfnamefont {T.}~\bibnamefont {Yang}}, \ and\ \bibinfo {author}
  {\bibfnamefont {B.}~\bibnamefont {Jia}},\ }\href@noop {} {\bibfield
  {journal} {\bibinfo  {journal} {Nature Communications}\ }\textbf {\bibinfo
  {volume} {11}},\ \bibinfo {pages} {1} (\bibinfo {year} {2020})}\BibitemShut
  {NoStop}%
\bibitem [{\citenamefont {Burger}\ \emph {et~al.}(2018)\citenamefont {Burger},
  \citenamefont {Fan}, \citenamefont {Lee}, \citenamefont {Forrest},\ and\
  \citenamefont {Lenert}}]{burger2018thin}%
  \BibitemOpen
  \bibfield  {author} {\bibinfo {author} {\bibfnamefont {T.}~\bibnamefont
  {Burger}}, \bibinfo {author} {\bibfnamefont {D.}~\bibnamefont {Fan}},
  \bibinfo {author} {\bibfnamefont {K.}~\bibnamefont {Lee}}, \bibinfo {author}
  {\bibfnamefont {S.~R.}\ \bibnamefont {Forrest}}, \ and\ \bibinfo {author}
  {\bibfnamefont {A.}~\bibnamefont {Lenert}},\ }\href@noop {} {\bibfield
  {journal} {\bibinfo  {journal} {ACS Photonics}\ }\textbf {\bibinfo {volume}
  {5}},\ \bibinfo {pages} {2748} (\bibinfo {year} {2018})}\BibitemShut
  {NoStop}%
\bibitem [{\citenamefont {Rogalski}(2005)}]{rogalski2005hgcdte}%
  \BibitemOpen
  \bibfield  {author} {\bibinfo {author} {\bibfnamefont {A.}~\bibnamefont
  {Rogalski}},\ }\href@noop {} {\bibfield  {journal} {\bibinfo  {journal}
  {Reports on Progress in Physics}\ }\textbf {\bibinfo {volume} {68}},\
  \bibinfo {pages} {2267} (\bibinfo {year} {2005})}\BibitemShut {NoStop}%
\bibitem [{\citenamefont {Hansen}\ and\ \citenamefont
  {Schmit}(1983)}]{hansen1983calculation}%
  \BibitemOpen
  \bibfield  {author} {\bibinfo {author} {\bibfnamefont {G.}~\bibnamefont
  {Hansen}}\ and\ \bibinfo {author} {\bibfnamefont {J.}~\bibnamefont
  {Schmit}},\ }\href@noop {} {\bibfield  {journal} {\bibinfo  {journal}
  {Journal of Applied Physics}\ }\textbf {\bibinfo {volume} {54}},\ \bibinfo
  {pages} {1639} (\bibinfo {year} {1983})}\BibitemShut {NoStop}%
\bibitem [{\citenamefont {Zhang}\ \emph {et~al.}(2020)\citenamefont {Zhang},
  \citenamefont {Du}, \citenamefont {Chen}, \citenamefont {Ang},\ and\
  \citenamefont {Ang}}]{zhang2020designing}%
  \BibitemOpen
  \bibfield  {author} {\bibinfo {author} {\bibfnamefont {X.}~\bibnamefont
  {Zhang}}, \bibinfo {author} {\bibfnamefont {J.}~\bibnamefont {Du}}, \bibinfo
  {author} {\bibfnamefont {J.}~\bibnamefont {Chen}}, \bibinfo {author}
  {\bibfnamefont {L.~K.}\ \bibnamefont {Ang}}, \ and\ \bibinfo {author}
  {\bibfnamefont {Y.~S.}\ \bibnamefont {Ang}},\ }\href@noop {} {\bibfield
  {journal} {\bibinfo  {journal} {Optics Letters}\ }\textbf {\bibinfo {volume}
  {45}},\ \bibinfo {pages} {5929} (\bibinfo {year} {2020})}\BibitemShut
  {NoStop}%
\bibitem [{\citenamefont {Berk}\ \emph {et~al.}(2006)\citenamefont {Berk},
  \citenamefont {Anderson}, \citenamefont {Acharya}, \citenamefont {Bernstein},
  \citenamefont {Muratov}, \citenamefont {Lee}, \citenamefont {Fox},
  \citenamefont {Adler-Golden}, \citenamefont {Chetwynd~Jr}, \citenamefont
  {Hoke} \emph {et~al.}}]{berk2006modtran5}%
  \BibitemOpen
  \bibfield  {author} {\bibinfo {author} {\bibfnamefont {A.}~\bibnamefont
  {Berk}}, \bibinfo {author} {\bibfnamefont {G.~P.}\ \bibnamefont {Anderson}},
  \bibinfo {author} {\bibfnamefont {P.~K.}\ \bibnamefont {Acharya}}, \bibinfo
  {author} {\bibfnamefont {L.~S.}\ \bibnamefont {Bernstein}}, \bibinfo {author}
  {\bibfnamefont {L.}~\bibnamefont {Muratov}}, \bibinfo {author} {\bibfnamefont
  {J.}~\bibnamefont {Lee}}, \bibinfo {author} {\bibfnamefont {M.}~\bibnamefont
  {Fox}}, \bibinfo {author} {\bibfnamefont {S.~M.}\ \bibnamefont
  {Adler-Golden}}, \bibinfo {author} {\bibfnamefont {J.~H.}\ \bibnamefont
  {Chetwynd~Jr}}, \bibinfo {author} {\bibfnamefont {M.~L.}\ \bibnamefont
  {Hoke}},  \emph {et~al.},\ }in\ \href@noop {} {\emph {\bibinfo {booktitle}
  {Algorithms and Technologies for Multispectral, Hyperspectral, and
  Ultraspectral Imagery XII}}},\ Vol.\ \bibinfo {volume} {6233}\ (\bibinfo
  {organization} {International Society for Optics and Photonics},\ \bibinfo
  {year} {2006})\ p.\ \bibinfo {pages} {62331F}\BibitemShut {NoStop}%
\bibitem [{\citenamefont {Chen}(2011)}]{chen2011theoretical}%
  \BibitemOpen
  \bibfield  {author} {\bibinfo {author} {\bibfnamefont {G.}~\bibnamefont
  {Chen}},\ }\href@noop {} {\bibfield  {journal} {\bibinfo  {journal} {Journal
  of Applied Physics}\ }\textbf {\bibinfo {volume} {109}},\ \bibinfo {pages}
  {104908} (\bibinfo {year} {2011})}\BibitemShut {NoStop}%
\bibitem [{\citenamefont {Bhatt}, \citenamefont {Kravchenko},\ and\
  \citenamefont {Gupta}(2020)}]{bhatt2020high}%
  \BibitemOpen
  \bibfield  {author} {\bibinfo {author} {\bibfnamefont {R.}~\bibnamefont
  {Bhatt}}, \bibinfo {author} {\bibfnamefont {I.}~\bibnamefont {Kravchenko}}, \
  and\ \bibinfo {author} {\bibfnamefont {M.}~\bibnamefont {Gupta}},\
  }\href@noop {} {\bibfield  {journal} {\bibinfo  {journal} {Solar Energy}\
  }\textbf {\bibinfo {volume} {197}},\ \bibinfo {pages} {538} (\bibinfo {year}
  {2020})}\BibitemShut {NoStop}%
\bibitem [{\citenamefont {Rouklove}(1966)}]{rouklove1966thermionic}%
  \BibitemOpen
  \bibfield  {author} {\bibinfo {author} {\bibfnamefont {P.}~\bibnamefont
  {Rouklove}},\ }\href@noop {} {\bibfield  {journal} {\bibinfo  {journal}
  {Space Programs;(United States)}\ }\textbf {\bibinfo {volume} {4}} (\bibinfo
  {year} {1966})}\BibitemShut {NoStop}%
\end{thebibliography}%

\end{document}